\documentclass[preprint2,10.5pt]{aastex}
\usepackage{epsfig}
\begin{document}
\title{\large{\rm{SPECTRAL TYPE AND RADIAL VELOCITY VARIATIONS IN THREE SRC VARIABLES}}}
\author{Kathleen E. Moncrieff$^1$, David G. Turner$^1$, C. Ian Short$^1$, Philip D. Bennett$^1$ \\ David D. Balam$^2$, Roger F. Griffin$^3$}
\affil{$^1$ Saint Mary's University, Halifax, Nova Scotia, Canada}
\affil{$^2$ Dominion Astrophysical Observatory, Victoria, British Columbia, Canada}
\affil{$^3$ The Observatories, Cambridge University, Cambridge, United Kingdom}
\email{\rm{kmoncrieff@ap.smu.ca}}

\begin{abstract}
SRC variables are M supergiants, precursors to Type II supernovae, that vary in brightness with moderately regular periods of order 100--1000 days. Although identified as pulsating stars that obey their own period-luminosity relation, few have been examined in enough detail to follow the temperature and spectral changes that they undergo during their long cycles. The present study examines such changes for several SRC variables revealed by CCD spectra obtained at the Dominion Astrophysical Observatory (DAO) during 2005--2009, as well as by archival spectra from the DAO (and elsewhere) for some stars from the 1960s to 1980s, and Cambridge radial velocity spectrometer measures for Betelgeuse. Described here is our classification procedure and information on the spectral type and radial velocity changes in three of the stars. The results provide insights into the pulsation mechanism in M supergiants.
\end{abstract}

\section{Introduction}

  Type C semiregulars, or SRC variables, constitute a portion of the ``forgotten'' family of late-type variable stars located in the M supergiant region of the Hertzsprung-Russell diagram. The more irregular LC variables are close cousins, and both types are poorly studied, mainly because of their lengthy primary cycles of $200-900$ days, although some also have superposed longer periods measuring thousands of days in duration.  They are evolved stars, most being M supergiants, with a few bright giants and a few late K-type stars thrown into the mix. They have masses of order $15-20 M_{\odot}$, effective temperatures below 4,000 K, and radii of at least several hundred $R_{\odot}$. Many are surrounded by large circumstellar dust shells created by their ongoing mass loss. SRC variables are also believed to be the precursors to most Type II supernovae.

  Recent summaries of the properties of SRC variables were presented by Turner (2006), Rohanizadegan et al.~(2006), and Turner et al.~(2006), the last two studies in connection with the luminous M3~Ia variable BC Cyg. The stars appear to obey their own period-luminosity relation (Turner et al.~ 2006). Spectroscopic studies of a large sample of M supergiants were made by Levesque et al.~(2005), including many of the SRC variables, and a study of period changes in BC Cyg reveals erratic changes of period in excess of what is typical of stars of late spectral type (Turner et al.~2009). It was demonstrated by Stothers (1969) forty years ago that pulsation can account for the primary periods of variability in such stars. But observational confirmation has been lacking, and detailed observational studies of the stars are few in number, mainly restricted to a limited spectroscopic and radial velocity survey by Joy (1942), the photometric survey of BC Cyg by Rohanizadegan et al.~(2006) and Turner et al.~(2006), and a brief survey of $\alpha$ Ori by Gray (2008). 

  In an attempt to remedy that situation, a spectroscopic survey of northern hemisphere SRC variables was initiated in 2005 using the 1.85m Plaskett telescope of the Dominion Astrophysical Observatory (DAO), Herzberg Institute of Astrophysics. Examination of archival photographic spectra for bright SRC variables, some from the David Dunlap Observatory (DDO), and scans of many of the original DAO plates have also been made, as well as a search for unpublished spectroscopic observations of the stars. Included are previously-unpublished radial velocity measures for Betelgeuse obtained by Griffin with the Cambridge radial velocity spectrometer. A comprehensive survey of the variables holds the promise of learning more about the basic properties of an intriguing group of massive stars, objects that may explode as supernovae within a few hundred years. This study examines changes in spectral type and radial velocity in three SRC variables: $\alpha$ Orionis, $\alpha$ Herculis, and S Persei. It is part of a larger project designed to improve our understanding of the pulsation mechanism in M supergiants and the evolutionary changes in SRC variables.
  
\section{Data Reduction}

  CCD spectra from the DAO were reduced with IRAF, first by applying bias corrections, trimming the spectra, and removing bad pixels with the {\it CCDPROC} package. Iron-argon arc spectra were used for wavelength calibration, and flat fielding was done with the {\it APFLATTEN} package. The {\it DOSLIT} package was used to extract one-dimensional (1-D) spectra from the two-dimensional (2-D) CCD output, as well as to wavelength-calibrate and dispersion-correct the former. The spectra were then normalized using IRAF's {\it CONTINUUM} package, which fits a polynomial to the data to find the continuum, and then outputs the ratio of the input spectra to the fitting function.

\section{Spectral Classification}

  Luminous M stars have been notoriously difficult to classify, primarily because most of the available spectral standards are SRC variables (Keenan \& McNeil~1976). A fairly straightforward scheme of temperature classification was set up by Keenan (see Gray \& Corbally~2009), based upon the visibility of specific band heads of TiO in blue-green spectra of the stars: $\lambda4954$ at M0, $\lambda4761$ at M1, $\lambda4804$ at M2, $\lambda4584$ at M3, $\lambda4626$ at M4, $\lambda4352$ at M5, $\lambda4395$ at M6, $\lambda4310$ at M7, and distinct TiO absorption longward of $\lambda4100$ at M8. That same scheme was employed here, along with the use of a VaO feature at $\lambda4389$ that is a useful luminosity discriminant for warmer M supergiants. Table~\ref{tab1} summarizes the lines used to determine spectral type and luminosity class for this study.

\begin{table}[h]
\centering
\begin{footnotesize}
\caption{Lines used for spectral classification.}
\label{tab1}
\vspace{2mm}
\begin{tabular}{lll}
\hline
\noalign{\smallskip}
$\lambda$ (\AA) &Species &Behavior\\
\noalign{\smallskip}
\hline
\noalign{\smallskip}
4861 &H$\beta$ &Appears to strengthen with decreasing \\
     &         &temperature when compared with \\
     &         &nearby continuum, which is suppressed \\
     &         &by the TiO band head.\\
\noalign{\smallskip}
\hline
\noalign{\smallskip}
4953 &TiO       &Strengthens with decreasing \\
     &band      &temperature. $\lambda 4953$ and other band \\
     &head      &heads are the primary determinants of \\
     &          &spectral type for M stars. The $\lambda4953$  \\
     &          &band head is easiest to use. \\
\noalign{\smallskip}
\hline
\noalign{\smallskip}
4389 &VaO       &Strengthens with increasing \\ 
     &blended   &luminosity when compared with \\
     &with      &nearby Fe I line at $\lambda4383$ \AA. \\
     &Fe I      &      \\
\noalign{\smallskip}
\hline
\end{tabular}
\end{footnotesize}
\end{table}
  
\subsection{Radial Velocity Determination}

  The IRAF package {\it RVIDLINES} was used to calculate radial velocities from the CCD and photographic spectra. {\it RVIDLINES} uses the differences between observed wavelengths in a spectrum and rest wavelengths of the same lines to calculate a radial velocity. The user creates a list of rest wavelengths, then marks several lines in the spectrum and enters the rest wavelengths of the marked lines. The software then uses the user-created list to identify as many more lines in the spectrum as possible, and computes a velocity from the average wavelength shift, along with an associated uncertainty. When the software is run in heliocentric mode, it uses the observatory location keyword in the header along with the date and time information to apply a heliocentric correction to the computed velocity automatically.
  
\section{Results}
\subsection{$\alpha$ Orionis}

\begin{table}[h]
\centering
\caption{$\alpha$ Orionis observations.}
\label{tab2}
\vspace{2mm}
\begin{footnotesize}
\begin{tabular}{cccll}
\hline
\noalign{\smallskip}
HJD &Phase &$V_r$ (km/s) &Sp. Type &Source\\
\noalign{\smallskip}
\hline
\noalign{\smallskip}
2433937 &0.4036 &$\cdots$ &M1.5-2 Iab &DDO \\
2437957 &0.9526 &$\cdots$ &M2 Iab &DDO \\
2439921 &0.6167 &$\cdots$ &M1.5-2 Iab &DDO \\
2441043 &0.2829 &$+19.9 \pm0.4$ &M1 Ia &DAO pg \\
2441132 &0.4943 &$+3.7 \pm0.1$ &M1 Ia &DAO pg \\
2441233 &0.7342 &$+10.3 \pm0.2$ &M2 Iab &DAO pg \\
2442458 &0.6439 &$-21.9 \pm0.8$ &M3 Ib &DAO pg \\
2442819 &0.5014 &$-22.6 \pm0.1$ &M3 Ib &DAO pg \\
2454041 &0.1562 &+23.5 &$\cdots$ &Griffin \\
2454057 &0.1942 &+23.8 &$\cdots$ &Griffin \\
2454079 &0.2464 &+23.3 &$\cdots$ &Griffin \\
2454124 &0.3530 &+22.6 &$\cdots$ &Griffin \\
2454186 &0.5023 &+18.5 &$\cdots$ &Griffin \\
2454360 &0.9140 &+18.7 &$\cdots$ &Griffin \\
2454391 &0.9888 &+22.3 &$\cdots$ &Griffin \\
2454414 &0.0421 &+24.7 &$\cdots$ &Griffin \\
2454421 &0.0587 &+24.9 &$\cdots$ &Griffin \\
2454429 &0.0797 &+24.7 &$\cdots$ &Griffin \\
2454443 &0.1110 &+24.7 &$\cdots$ &Griffin \\
2454472 &0.1797 &+22.2 &$\cdots$ &Griffin \\
2454490 &0.2246 &+21.8 &$\cdots$ &Griffin \\
2454508 &0.2673 &+21.4 &$\cdots$ &Griffin \\
2454523 &0.3029 &+21.0 &$\cdots$ &Griffin \\
2454757 &0.8571 &+22.8 &$\cdots$ &Griffin \\
2454779 &0.9091 &+25.1 &$\cdots$ &Griffin \\
2454804 &0.9683 &+25.5 &$\cdots$ &Griffin \\
2454845 &0.0678 &+26.6 &$\cdots$ &Griffin \\
2454876 &0.1415 &+26.9 &$\cdots$ &Griffin \\
2454897 &0.1911 &+26.3 &$\cdots$ &Griffin \\
2454929 &0.2670 &+25.4 &$\cdots$ &Griffin \\
2455114 &0.7050 &+25.1 &$\cdots$ &Griffin \\
2455128 &0.7383 &+25.2 &$\cdots$ &Griffin \\
2455167 &0.8307 &+23.9 &$\cdots$ &Griffin \\
2455187 &0.8780 &+23.4 &$\cdots$ &Griffin \\
2455214 &0.9443 &+23.2 &$\cdots$ &Griffin \\
2455227 &0.9729 &+23.3 &$\cdots$ &Griffin \\
2455258 &0.0487 &+25.2 &$\cdots$ &Griffin \\
2455295 &0.1364 &+24.6 &$\cdots$ &Griffin \\
\noalign{\smallskip}
\hline
\end{tabular}
\end{footnotesize}
\end{table}

\begin{figure}[h]
\centerline{\epsfig{file=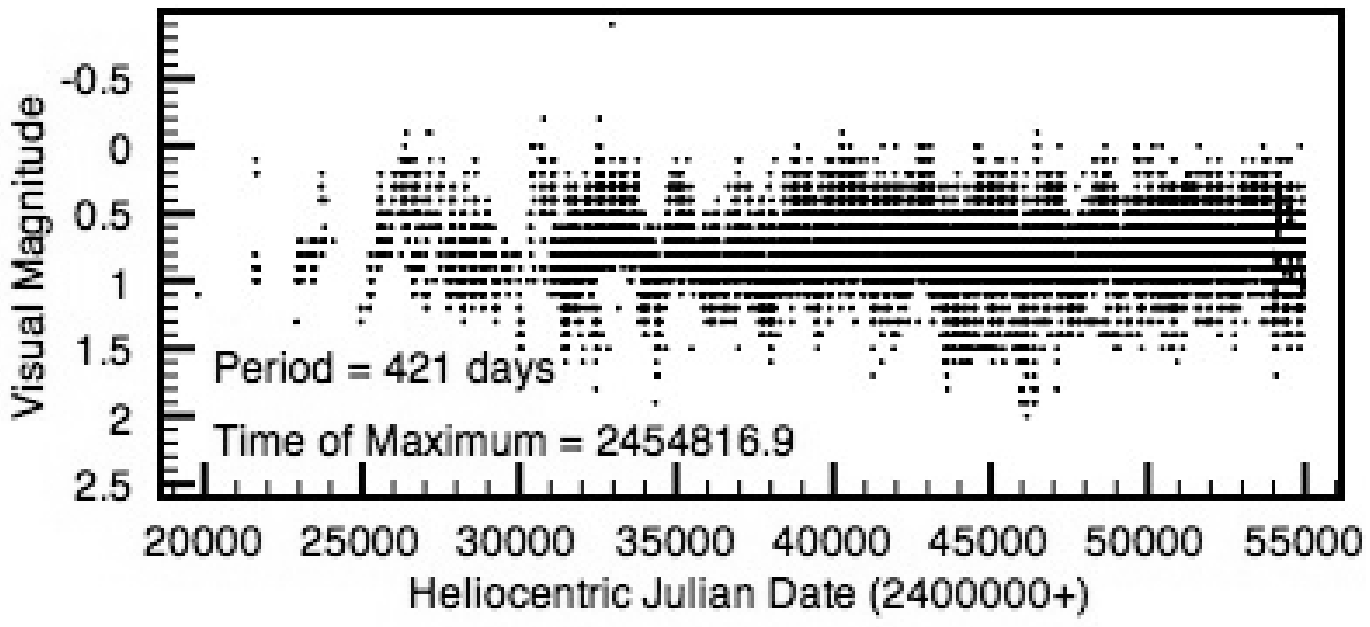, scale=0.50}}
\centerline{\epsfig{file=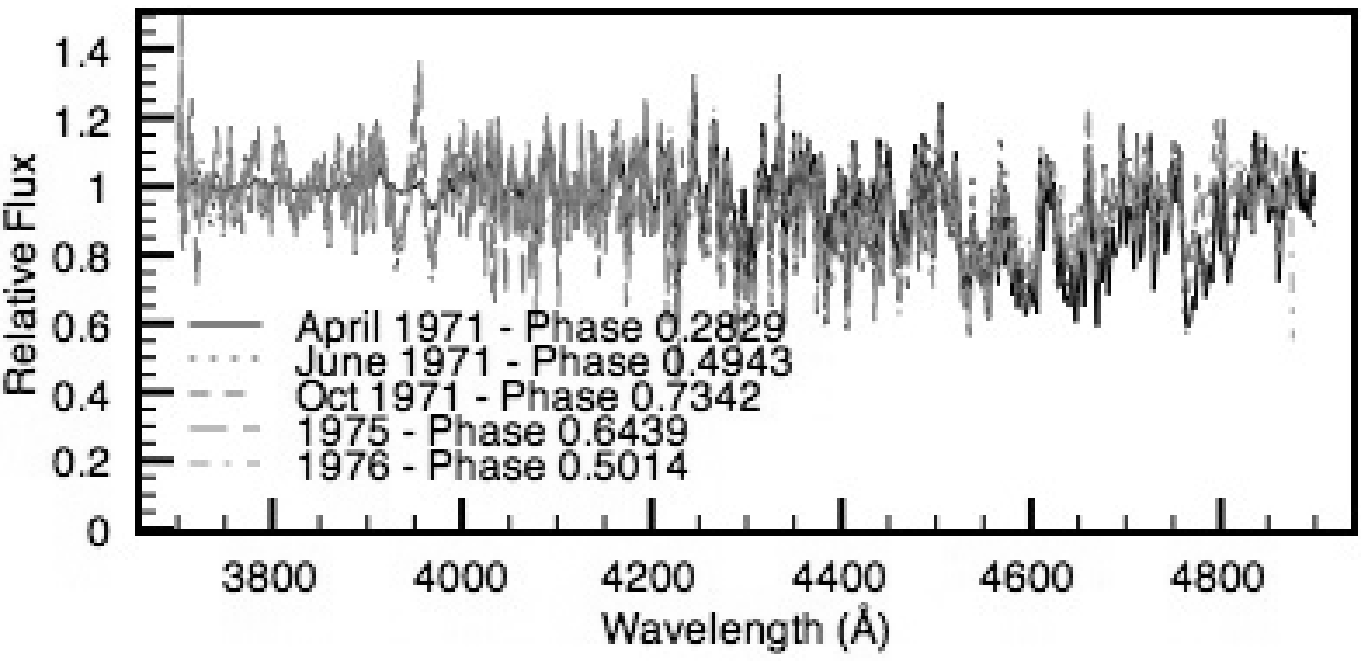, scale=0.50}}
\centerline{\epsfig{file=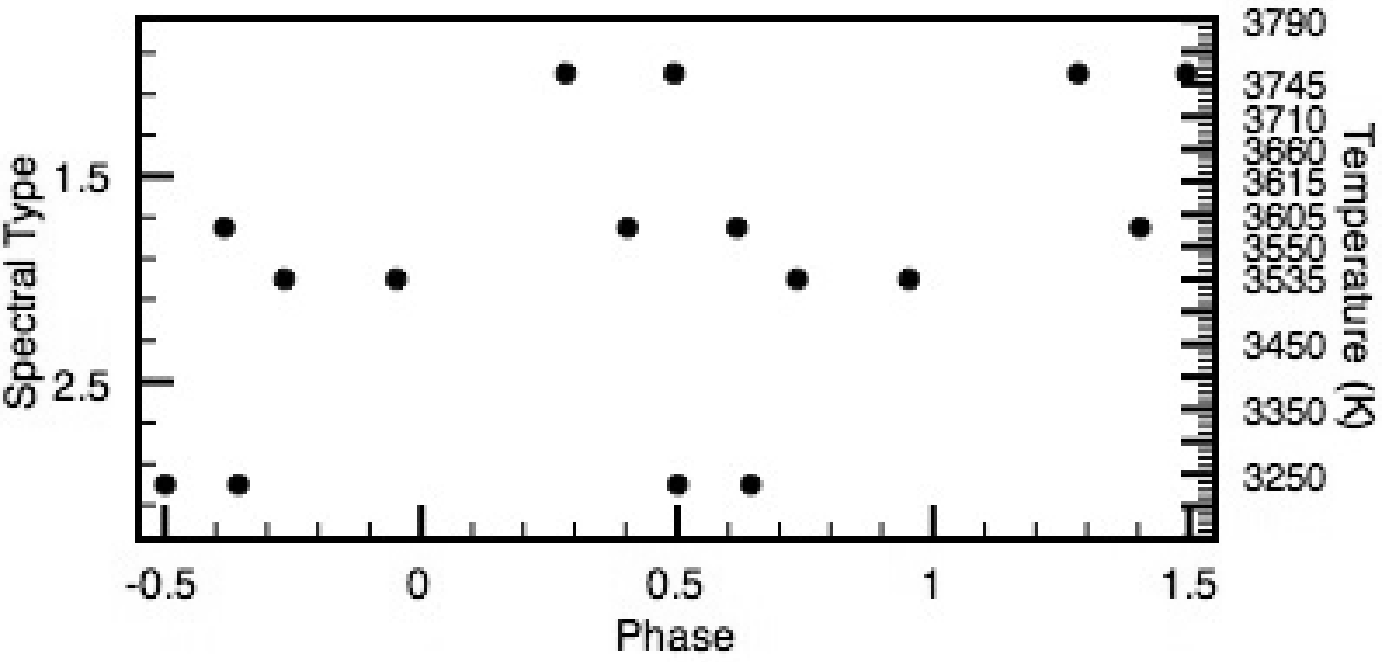, scale=0.50}}
\centerline{\epsfig{file=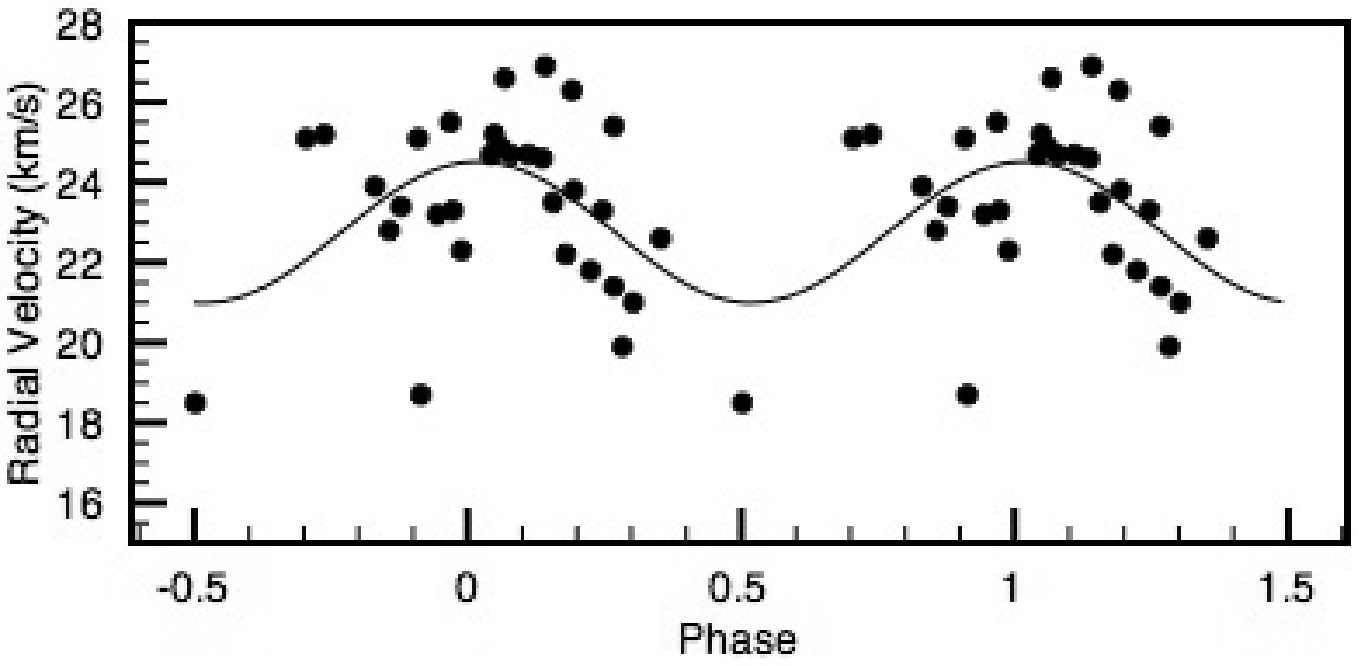, scale=0.50}}
\caption{\small{$\alpha$ Ori data. From top: visual light curve from AAVSO measurements, recent spectra, spectral types and temperatures as functions of phase, and radial velocity as a function of phase.}}
\label{fig1}
\end{figure}

\begin{figure}[h]
\centerline{\epsfig{file=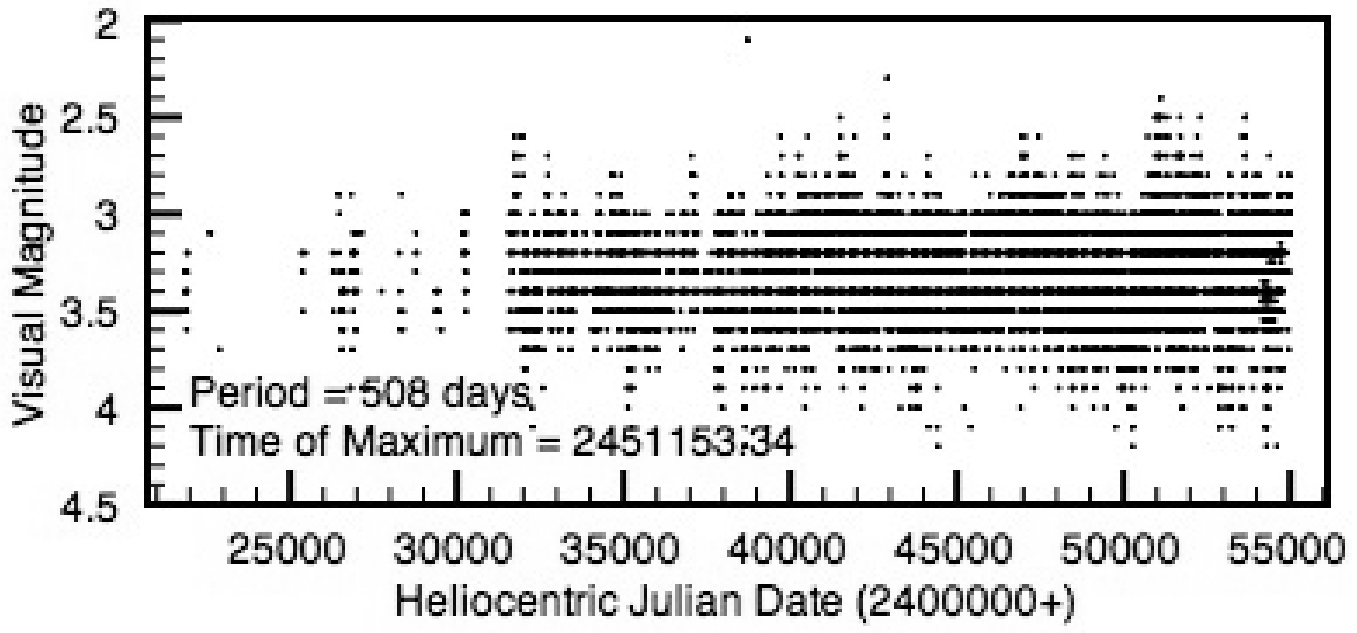, scale=0.50}}
\centerline{\epsfig{file=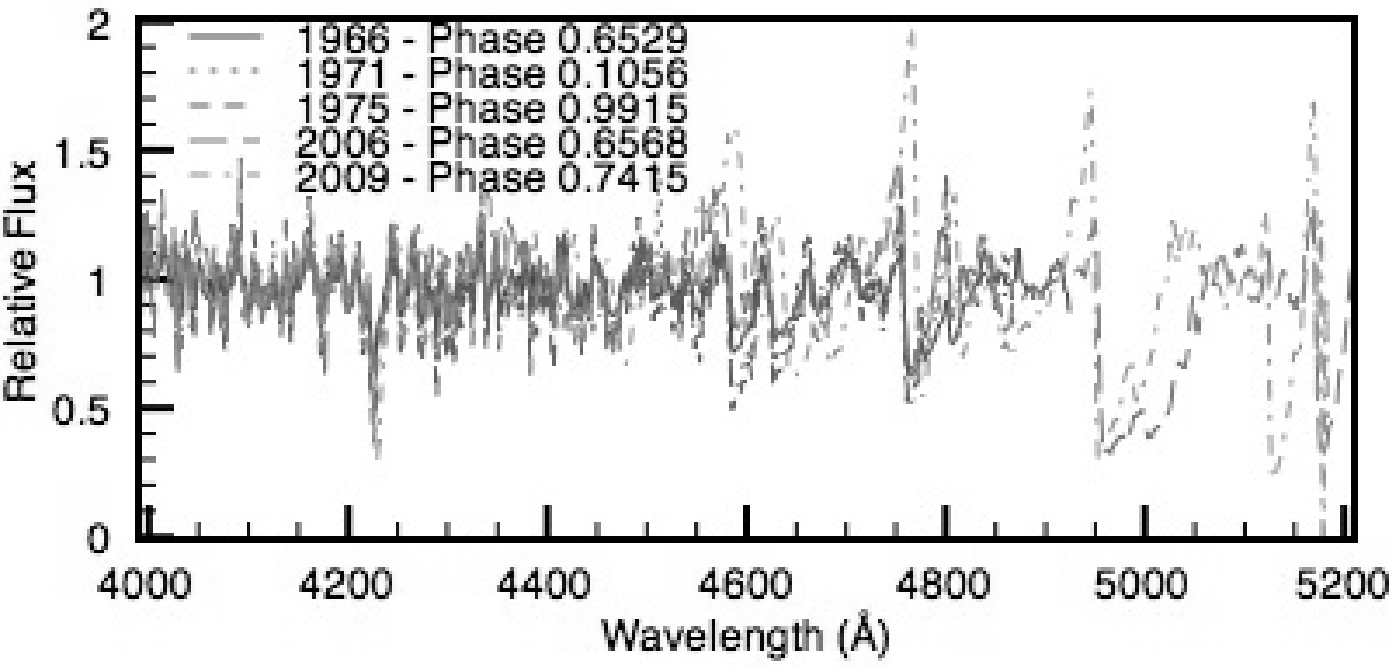, scale=0.50}}
\centerline{\epsfig{file=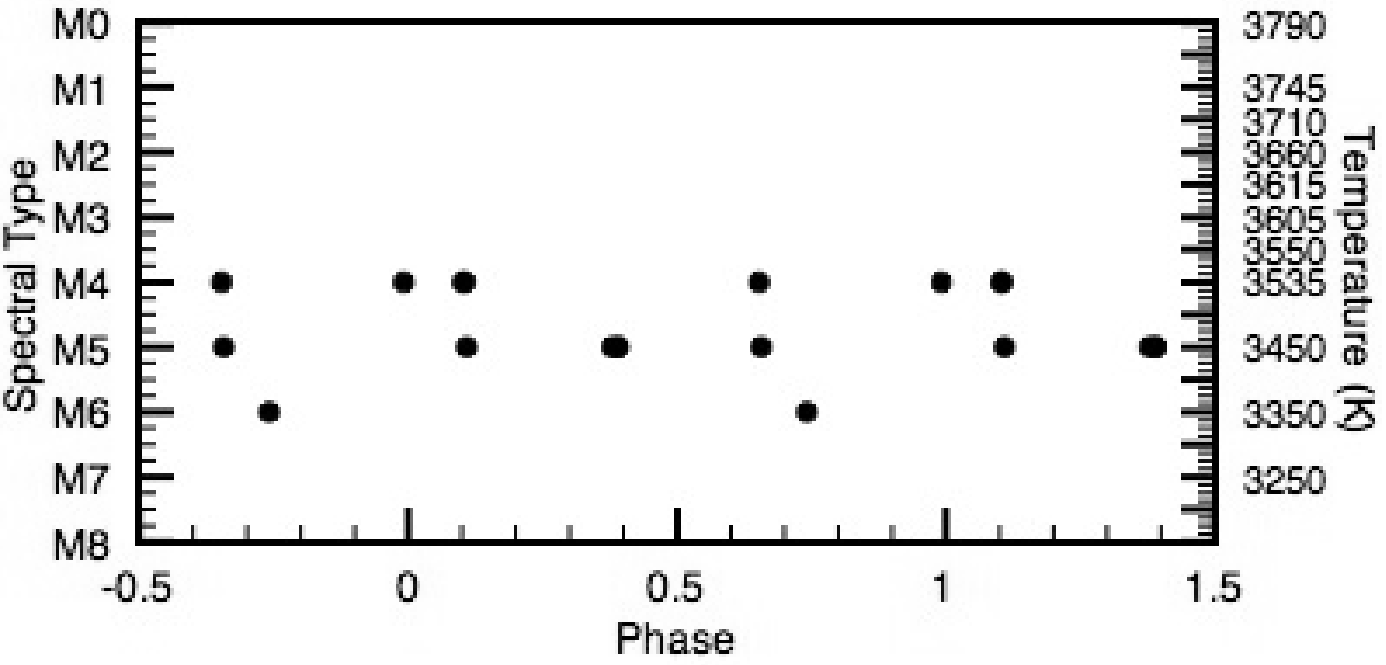, scale=0.50}}
\centerline{\epsfig{file=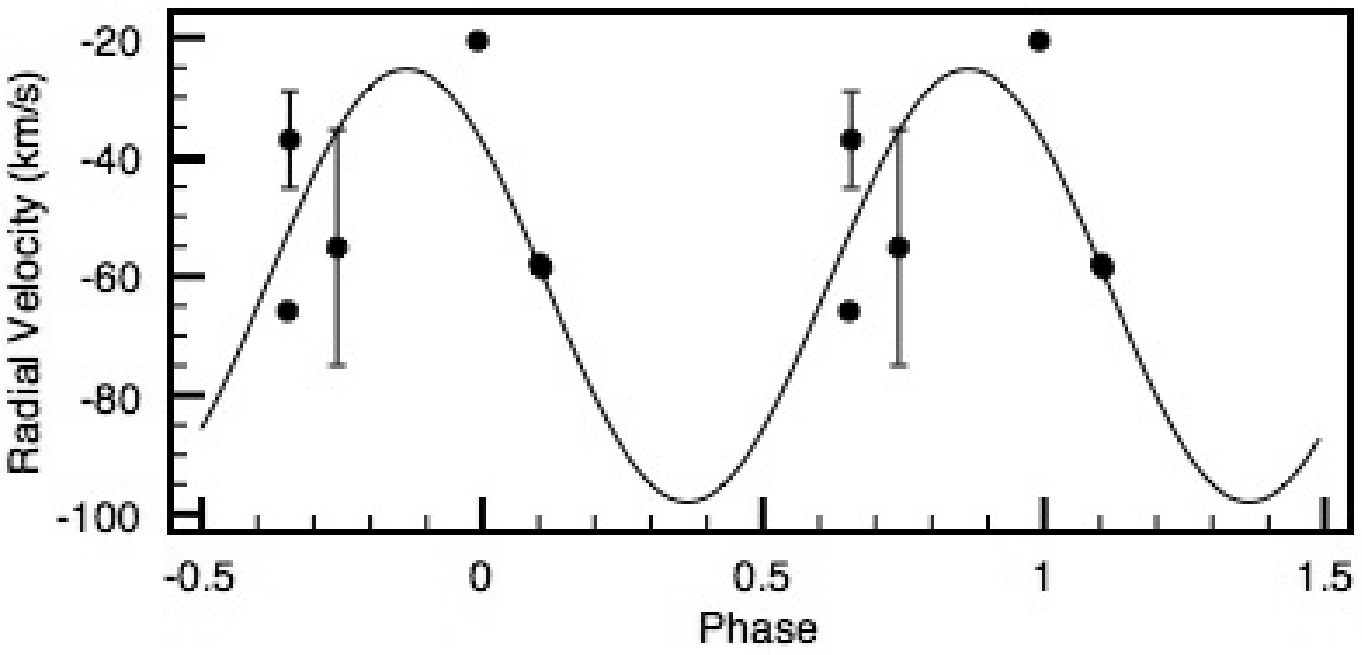, scale=0.50}}
\caption{\small{$\alpha$ Her data. From top: visual light curve from AAVSO measurements, recent spectra, spectral types and temperatures as functions of phase, and radial velocity as a function of phase.}}
\label{fig2}
\end{figure}

\begin{figure}[h]
\centerline{\epsfig{file=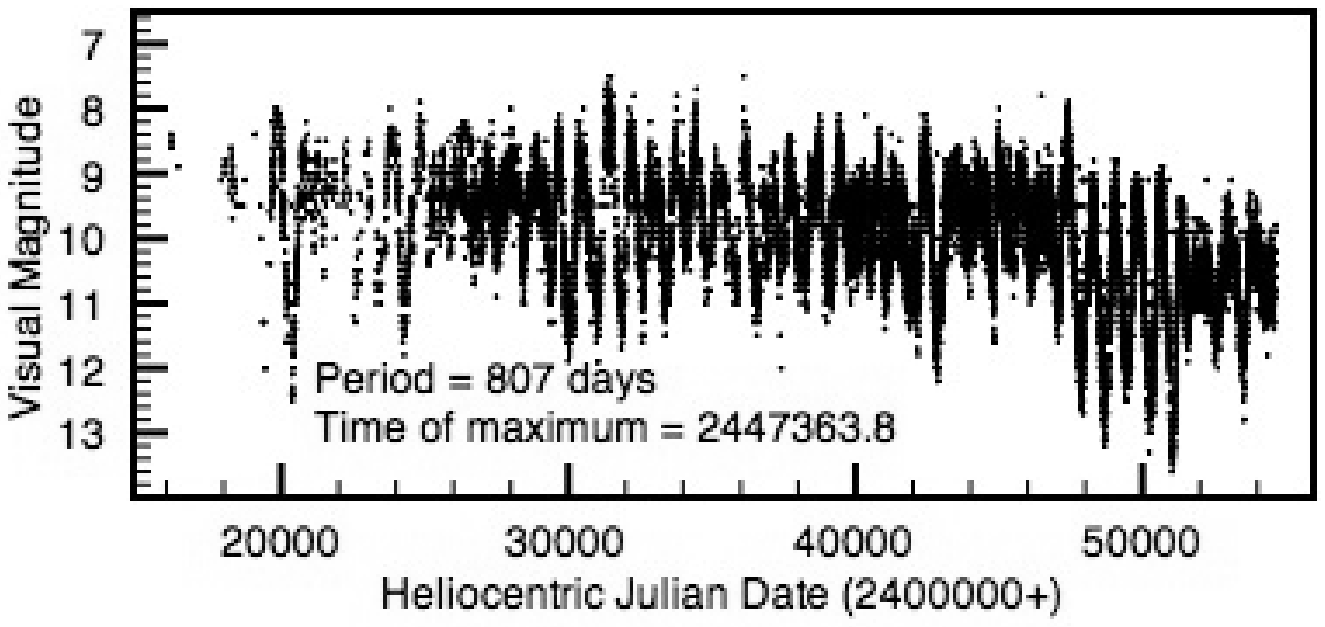, scale=0.50}}
\centerline{\epsfig{file=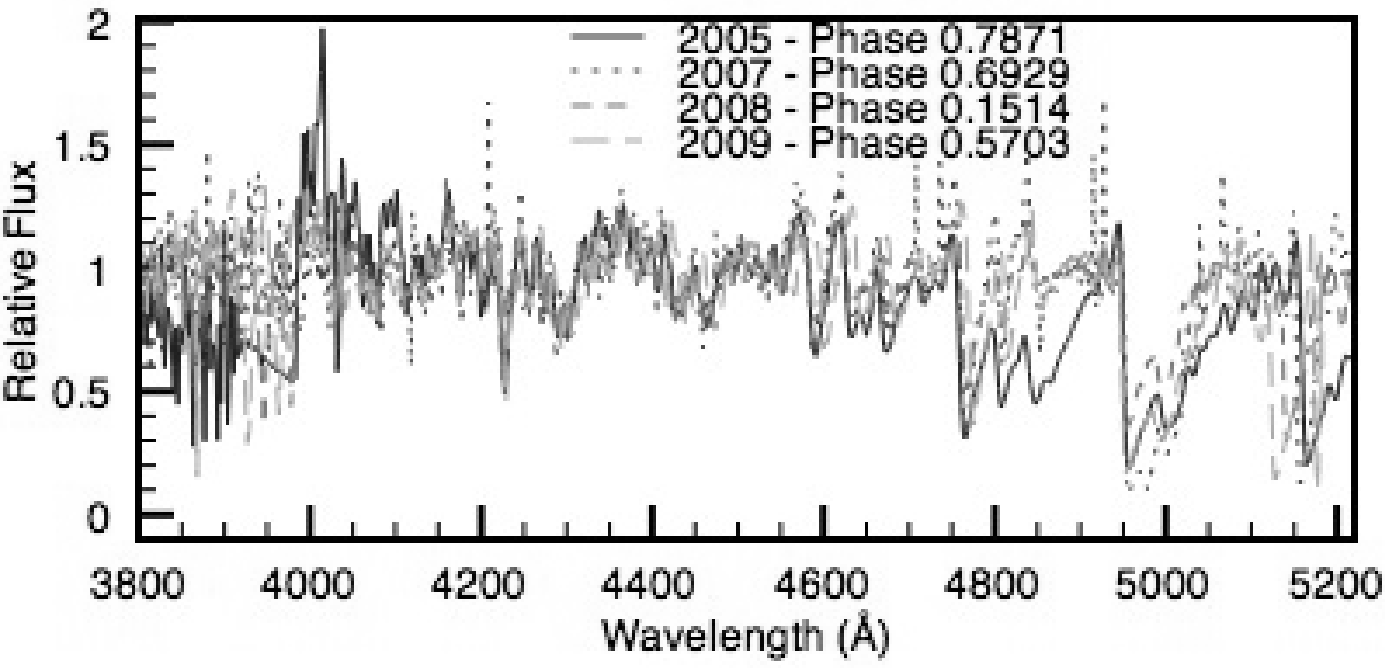, scale=0.50}}
\centerline{\epsfig{file=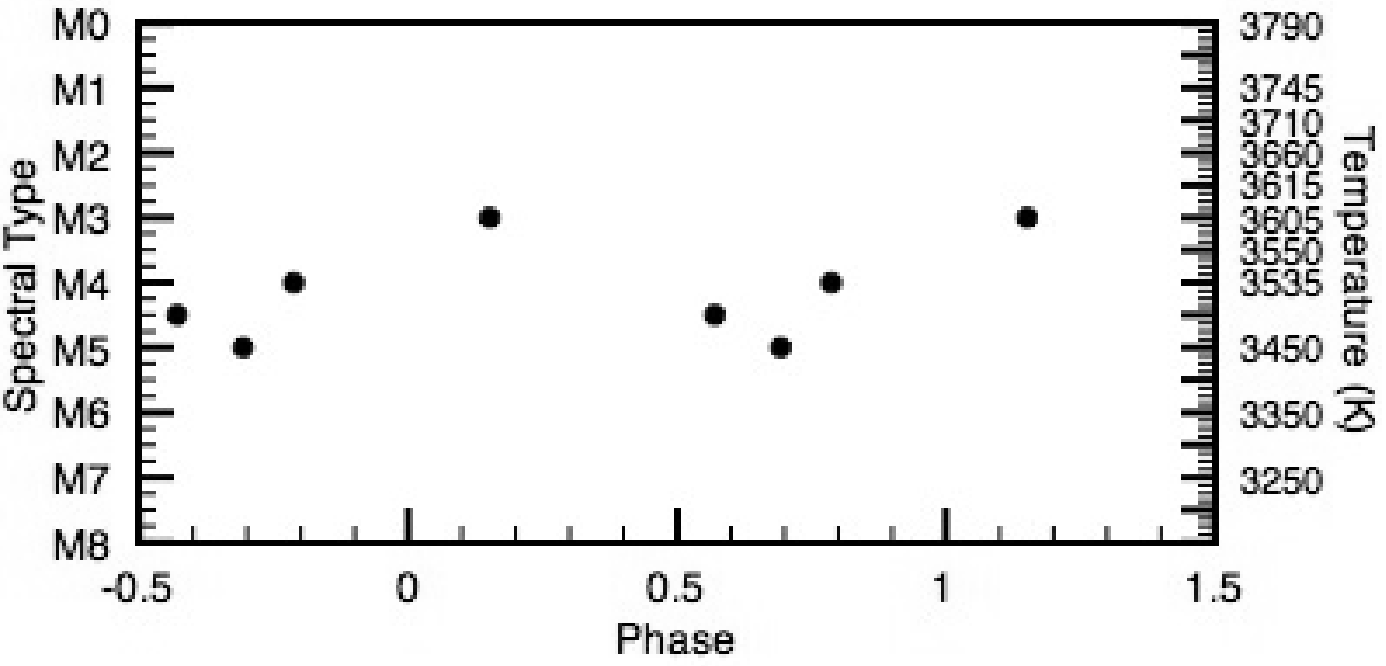, scale=0.50}}
\centerline{\epsfig{file=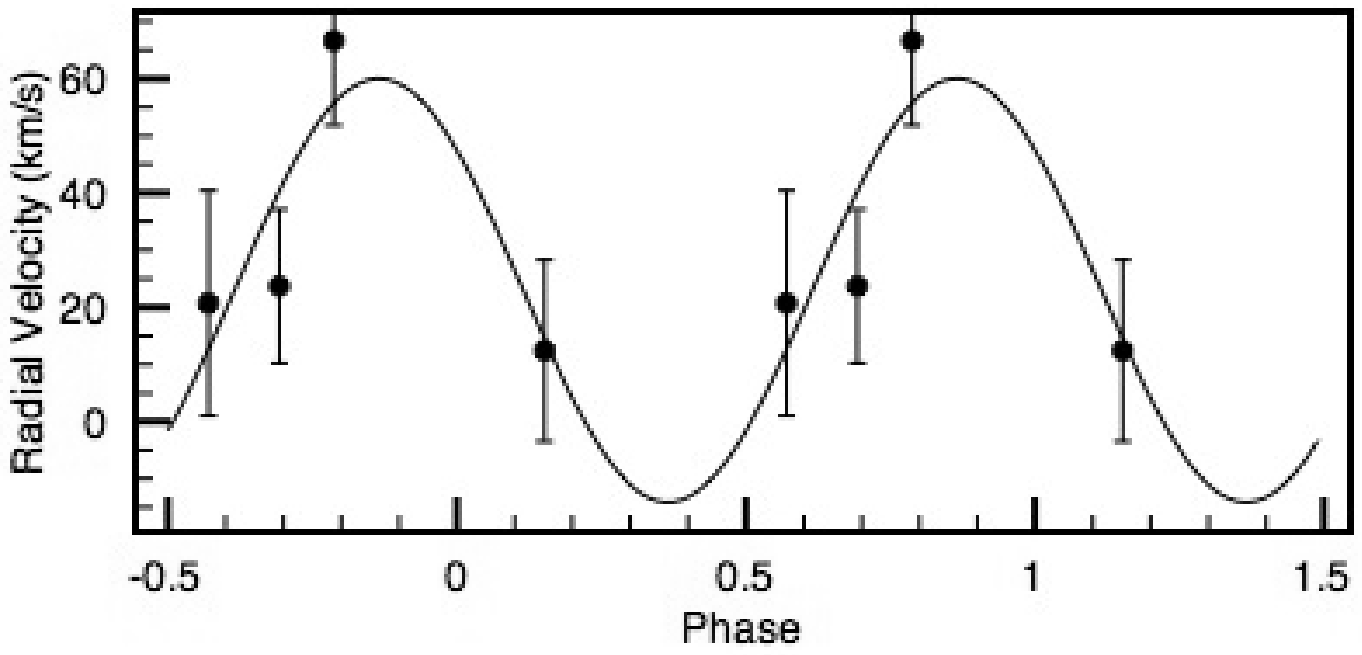, scale=0.50}}
\caption{\small{S Per data. From top: visual light curve from AAVSO measurements, recent spectra, spectral types and temperatures as functions of phase, and radial velocity as a function of phase.}}
\label{fig3}
\end{figure}

  A comparison of AAVSO photometric information on the star with newly-derived and available spectral type and radial velocity data indicates that $\alpha$ Ori reaches its latest spectral type and lowest luminosity near light minimum. Radial velocity maximum (greatest photospheric recession) is reached near light maximum and the star reaches mean radial velocity (indicative of smallest dimensions) near light phase 0.27. Table~\ref{tab2} summarizes the spectroscopic results and Fig.~\ref{fig1} displays the available data for $\alpha$ Ori.\\[2mm]
  
\subsection{$\alpha$ Herculis}
 
\begin{table}[h]
\centering
\caption{$\alpha$ Herculis observations.}
\label{tab3}
\vspace{2mm}
\begin{footnotesize}
\begin{tabular}{cccll}
\hline
\noalign{\smallskip}
HJD &Phase &$V_r$ (km/s) &Sp. Type &Source\\
\noalign{\smallskip}
\hline
\noalign{\smallskip}
2427978 &0.3787 &$\cdots$ &M5 Iab-Ib &DDO \\
2427984 &0.3905 &$\cdots$ &M5 Iab-Ib &DDO \\
2439293 &0.6529 &$-65.9 \pm0.7$ &M4 Ia &DAO pg \\
2440033 &0.1089 &$\cdots$ &M5 Ib &DDO \\
2441047 &0.1056 &$-58.7 \pm0.4$ &M4 Ia &DAO pg \\
2442513 &0.9915 &$-20.5 \pm0.3$ &M4 Ia &DAO pg \\
2454027 &0.6568 &$-37.1 \pm8.1$ &M5 Iab &DAO CCD \\
2455086 &0.7415 &$-55.2 \pm19.8$ &M6 Ib &DAO CCD \\
\noalign{\smallskip}
\hline
\end{tabular}
\end{footnotesize}
\end{table}

  A comparison of AAVSO photometric information on $\alpha$ Her with newly-derived and available spectral type and radial velocity data indicates that it reaches smallest dimensions at phase 0.12 following light maximum. There is scatter in its spectral classifications, yet it appears to reach greatest luminosity at earliest spectral type. The star belongs to an optical triple. Table~\ref{tab3} summarizes the spectroscopic results and Fig.~\ref{fig2} displays the available data for $\alpha$ Her.

\subsection{S Persei}
  
\begin{table}[h]
\centering
\caption{S Persei observations.}
\label{tab4}
\vspace{2mm}
\begin{footnotesize}
\begin{tabular}{cccll}
\hline
\noalign{\smallskip}
HJD &Phase &$V_r$ (km/s) &Sp. Type &Source\\
\noalign{\smallskip}
\hline
\noalign{\smallskip}
2453648 &0.7871 &$66.6 \pm14.8$ &M4 Ia &DAO CCD \\
2454379 &0.6929 &$23.7 \pm13.5$ &M5 Iab &DAO CCD \\
2454749 &0.1514 &$12.4 \pm15.9$ &M3 Ia+ &DAO CCD \\
2455087 &0.5703 &$20.7 \pm19.7$ &M4.5 Ia &DAOCCD \\
\noalign{\smallskip}
\hline
\end{tabular}
\end{footnotesize}
\end{table}

  A comparison of AAVSO photometric information on S Per with newly-derived spectral type and radial velocity data indicates that it also reaches smallest dimensions at phase 0.11 following light maximum, and appears to reach greatest luminosity and earliest spectral type near light maximum.  Table~\ref{tab4} summarizes the spectroscopic results and Fig.~\ref{fig3} displays the available data for S Per.

\section{Conclusions}

  All three survey stars appear to reach their smallest dimensions following light maximum at photometric phases 0.11--0.27. The photospheric temperature changes follow the changes in overall size, with $\alpha$ Ori and S Per reaching their coolest photospheric temperatures near light minimum and greatest atmospheric extent. Like S Per, $\alpha$ Ori reaches its highest photospheric temperatures just prior to reaching its smallest overall dimensions. Scatter in the spectral classifications for $\alpha$ Her make it difficult to determine exactly when it reaches hottest photospheric temperature, but in all three stars greatest luminosity appears to be coincident with such phases, implying that the $T_{\rm eff}^4$ term dominates the luminosity equation. Similar analyses are currently being performed on $\sim20$ additional SRC variables, which will permit us to determine if other members of the class display the same general characteristics.

\subsection*{\sc{acknowledgements}}
\small{We are grateful to the DAO for allocating time on the 1.85-m telescope for this project. We also acknowledge with thanks the variable star observations from the American Association of Variable Star Observers (AAVSO) International Database contributed by observers worldwide and used in this research. This research used the POLLUX database (http://pollux.graal.univ- montp2.fr), operated at GRAAL (Universit\'e Montpellier II - CNRS, France) with the support of the PNPS and INSU.}

\end{document}